\documentclass{article}

\usepackage{spconf,amsmath,epsfig}
\usepackage{hyperref}
\usepackage{amsmath}
\usepackage{cite}
\usepackage{amssymb}
\usepackage{algorithm}
\usepackage{algpseudocode}
\usepackage{balance}
\usepackage{bbm}
\usepackage{rotating}

\usepackage{hyperref}
\hypersetup{
    bookmarks=true,         % show bookmarks bar?
    unicode=false,          % non-Latin characters in Acrobats bookmarks
    pdftoolbar=true,        % show Acrobats toolbar?
    pdfmenubar=true,        % show Acrobats menu?
    pdffitwindow=false,     % window fit to page when opened
    pdfstartview={FitH},    % fits the width of the page to the window
    pdftitle={CPCA},    % title
    pdfauthor={Camps-Valls},     % author
    pdfsubject={CPCA},   % subject of the document
    pdfcreator={Camps-Valls},   % creator of the document
    pdfproducer={Camps-Valls}, % producer of the document
    pdfkeywords={keyword1} {key2} {key3}, % list of keywords
    pdfnewwindow=true,      % links in new window
    colorlinks=true,       % false: boxed links; true: colored links
    linkcolor=blue,          % color of internal links (change box color with linkbordercolor)
    citecolor=blue,        % color of links to bibliography
    filecolor=blue,      % color of file links
    urlcolor=blue           % color of external links
}

\def\x{{\mathbf x}}

\def\k{{\mathbf k}}

\def\C{{\mathbf C}}
\def\X{{\mathbf X}}
\def\G{{\mathbf G}}
\def\Xt{\tilde{\mathbf X}}
\def\Pt{\tilde{\boldsymbol{\Phi}}}
\def\K{{\mathbf K}}

\def\Real{{\mathbbm{R}}}
\def\Comp{{\mathbbm{C}}}
\newcommand{\imag}{\mathbbm{i}}
\newcommand{\HH}{\mathbbm{H}}
\usepackage{color}

\onecolumn

\title{Nonlinear Complex PCA\\for spatio-temporal analysis of global soil moisture}

\name{Diego Bueso, Maria Piles, Gustau Camps-Valls
\thanks{ \copyright IEEE. Personal use of this material is permitted.  Permission from IEEE must be obtained for all other users,including reprinting/republishing this material for advertising or promotional purposes, creating new collectiveworks for resale or redistribution to servers or lists, or reuse of any copyrighted components of this work in otherworks.  DOI: 10.1109/IGARSS.2018.8518155}
	\thanks{This work is partly funded by the European Research Council (ERC) under the ERC-CoG-2014 SEDAL project (grant agreement 647423).}
}
\address{Image Processing Lab (IPL), Universitat de Val\`encia, Val\`encia, Spain}

\begin{document}
%\ninept

\maketitle

\begin{abstract}
Soil moisture (SM) is a key state variable of the hydrological cycle, needed to monitor the effects of a changing climate on natural resources. Soil moisture is highly variable in space and time, presenting seasonalities, anomalies and long-term trends, but also, and important nonlinear behaviours. Here, we introduce a novel fast and nonlinear complex PCA method to analyze the spatio-temporal patterns of the Earth's surface SM. We use global SM estimates acquired during the period 2010-2017 by ESA's SMOS mission. Our approach unveils both time and space modes, trends and periodicities unlike standard PCA decompositions. 
Results show the distribution of the total SM variance among its different components, and indicate the dominant modes of temporal variability in surface soil moisture for different regions. The relationship of the derived SM spatio-temporal patterns with El Ni{\~n}o Southern Oscillation (ENSO) conditions is also explored. 
%Results show the derived SM components and their association with El Ni{\~n}o Southern Oscillation (ENSO). 

\end{abstract}


%%%%%%%%%%%%%%%%%%%%%%%%%%%%%%%%%%%%%%%%%%%%%%%%%%%%%%%%%%%%%%%%%%%%%%%%%%%%%%%%%%%%%%%%%%%%%%%%%%
\section{Introduction}
\label{sec:intro}

Soil moisture exerts a strong control not only on the water cycle but also in ecosystem functioning. In this regard, monitoring changes in soil moisture is critical, since they have a direct impact on agricultural productivity, forestry, and ecosystem health ~\cite{seneviratne2010}. In this work, we present a novel methodology ,the rotated kernel complex Principal component analysis (ROCK PCA) that allows characterizing global SM spatio-temporal dynamics. To do so, we use the first seven years (2010-2017) of soil moisture estimates obtained from the ESA-led Soil Moisture and Ocean Salinity (SMOS) mission.

Principal component analysis, also known in geophysics as Empirical Orthogonal Functions (EOFs), is widely used to obtain  compact representations of the signal~\cite{Arenas13}, and has been widely exploited to obtain spatio-temporal features in climatological studies~\cite{bauer2013,volkov2014,forootan2016}. However, PCA is computationally demanding with a large number of features, and at the same time the spatio-temporal components are not disentangled in the feature space. In this work, we propose a method that alleviates both problems. The complex PCA is first applied to spatial and temporal components through the Hilbert transform, thus leading to spatial and temporally explicit eigendecompositions easy to analyze. It exploits the fact that the eigendecomposition of the covariance and the Gram matrix return the same results, and hence the computational cost can be drastically reduced from quadratic ${\mathcal O}(n^2 t)$  to linear in the number of pixels ${\mathcal O}(t^2 n)$~\cite{sharma2007}. The use of the Gram matrix is not incidental, and allows further improvements, such as the direct kernelization of the method~\cite{CampsValls09}, thus allowing to derive a fast nonlinear spatio-temporal PCA. An extra rotation, the promax oblique method, is also introduced for extra versatily as it allows us to find the optimal parameters for the decomposition and make the principal components physically interpretable.

The rest of the paper is outlined as follows. Section 2 introduces our method for spatio-temporal data decomposition: we first fix the notation, review the standard approaches in the literature, and introduce our nonlinear (kernel-based) complex PCA and the rotation approach. Section 3 presents the experimental results on SMOS data to uncover explicit temporal and spatial patterns of soil moisture, which are then related to ENSO events. Section 4 concludes with some remarks and future work.

%%%%%%%%%%%%%%%%%%%%%%%%%%%%%%%%%%%%%%%%%%%%%%%%%%%%%%%%%%%%%%%%%%%%%%%%%%%%%%%%%%%%%%%%%%%%%%%%%%
\section{Nonlinear spatio-temporal data analysis}

\subsection{Notation}

Let us define a spatio-temporal data cube $\C\in\Real^{r\times c \times t}$, defined in a $r\times c$ spatial grid and $t$ temporal observations. The cube can be reshaped into matrix form as $\Xt=[x_1 , \cdots ,x_n]\in\Real^{t\times n}$, where the tilde indicates the column-wise centering operation for each time series $\x_i \in \Real^{t \times 1},i=1,\cdots ,n$. A standard way to analyze the feature relations contained in the data is to diagonalize its covariance matrix, $\C = \frac{1}{t}\Xt^\top\Xt \in\Real^{n\times n}$. However, given the high number of pixel observations, $n$, obtaining the eigenvalues and eigenvectors involves a high computational cost. A simple efficient alternative is to eigedecompose the Gram matrix: $\G = \Xt\Xt^\top \in\Real^{t\times t}$, which returns exactly the same solution up to a scaling factor. This is known as the primal-dual solution of the PCA.

\subsection{Complex PCA}

The main problem by following the standard PCA approach, is that interpretability of the components is hardly accessible. Eigenvectors and eigenvalues do not longer have a clear, physically meaningful interpretation in terms of spatial and temporal coordinates in the projection space~\cite{koscielny1982}. The decomposition of the datacube into spatial and temporal eigecomponents is a traditional methodology commonly known as {\em complex} PCA~\cite{complexPCA}. The complex modality of PCA returns a more accurate decomposition and interpretable eigenvectors for geophysical data analysis than the plain PCA version. %, where the complex output data allows to applicate complex signal analysis method to the eigendecompose. 
Formally, the complex PCA essentially applies the Hilbert transform to a signal $\x_{i}(t)$:
$$
H(x_{i}(t)) = 
\frac{1}{\pi} \int_{-\infty}^{+\infty} \dfrac{x_{i}(\tau)}{t-\tau} \,d\tau,i=1,\cdots ,n.
$$
The Hilbert-transformed point is now expressed as $\x_{i,h} = \x_{i}+\imag H(\x_{i})$, and hence the centered Hilbert-transformed data matrix becomes:
$$\Xt_h = \Xt + \imag H(\Xt),$$
where we define the Hermitian of $\X_h\in\Comp$ as $\X_h^\HH\in\Comp$, and orthogonality holds, $\x_i \perp H(\x_i)$. Now, one can easily demonstrate that the covariance matrix of Hilbert-transformed data reduces to
$$\C_h = \frac{1}{t}\Xt_h^\HH\Xt_h = \C_{R} + \imag \C_{I} \in \Comp^{n\times n},$$
and the corresponding, and more convenient, centered Gram matrix is: 
$$\G_h = \Xt_h\Xt_h^\HH = \G_{R} + \imag \G_{I}\in\Comp^{t\times t},$$
which is then used for the sake of a fast eigendecomposition. Note that now we retrieve separate spatial and temporal eigenvalues and eigenvectors, which allows us to study the signal in those terms. Making the eigendecomposition of the Gram matrix (classically, the covariance matrix), we obtain the eigenvalues $(\boldsymbol{\lambda}_{h} \in \Real^{t\times 1})$ which are the explained variance for each Principal Component (PC) and the eigenvectors or PCs $(\boldsymbol{V}_{h}\in \Comp^{t\times t})$ which in Gram matrix case will be the estimate time series. The spatial features will be the projection onto the PCs $(\boldsymbol{Xp}_{h}=\boldsymbol{V}_{h}^{\HH}\Xt_{h} \in \Comp^{t\times n}).$

\subsection{Kernelized Complex PCA}

Working with Gram matrices instead of covariances allows us to directly derive a nonlinear version of the previous complex PCA method by means of the kernel trick~\cite{CampsValls09,Arenas13}. Working with a nonlinear version can avoid adopting orthogonality constraints~\cite{koscielny1982}, which has a traditional solution with the rotated PCA~\cite{rotatedPCA}. The kernel PCA~\cite{Arenas13} could also achieve good representation of the data features but the spatial and temporal components would still become intertwined. %in s eigenvectors, we will make a simple demonstration in section ~\ref{subsec:example}. 

Alternatively, we derive here a kernelized version of the complex PCA. Let us define a feature map into a Hilbert space, $\boldsymbol{\phi}: \x_i\mapsto \boldsymbol{\phi}(\x_i)\in{\mathcal H}$, which is endorsed with a dot, product called {\em kernel function}, $k(\x_i,\x_j) = \boldsymbol{\phi}(\x_i)^\top\boldsymbol{\phi}(\x_j)\in\Real$. The (centered) kernel matrix groups all dot products into a matrix defined as $\K = \Pt\Pt^\top \in\Real^{t\times t}$. The kernel function essentially computes similarities between feature vectors. Similarly, a kernel feature vector contains all similarities between a test point $\x_*$ and all the points in the datasets, and is defined as $\k_*:=[k(\x_*,\x_1),\ldots,k(\x_*,\x_n)]^\top\in\Real^{n\times 1}$. Then, it simply follows from the application of the Hilbert transform and the orthogonality property that the corresponding kernelized complex PCA reduces to eigendecompose 
$$\K_h = \Pt_h\Pt_h^\HH = \K_{R} + \imag \K_{I} \in\Comp^{t\times t},$$
and thus we can analyze the signal in nonlinear terms, and separate components in space and time as a classic kernel PCA ~\cite{muller1998}. Taking into account the circularity of our complex data from Hilbert transform we use a complex RBF (Radial Basis Function) kernel function  introduced by ~\cite{bouboulis2011} defined as $k(\x_{i},\x_{j})= exp(-(\x_{i}-\x_{j}^*)(\x_{i}-\x_{j}^*)^T/2\sigma^2)$ which hold the circular properties of input data and is defined as Hermitian and semi-positive definite. Our nonlinear decomposition is defined by the kernel parameter $\sigma$ and the number of PCs. This reduces our problem to compute the eigendecomposition of the kernel matrix $\K$ to  obtain $\boldsymbol{V}_{h}$. In the next section we show the way to fit the $\sigma$ hyperparameter using a rotated approach while making an oblique rotation of the PCs to create a more amenable physically-meaningful representation of data. 

\subsection{Promax Rotation}

Using Kernel functions allows to finding nonlinear representations of the data, but frequently cannot tackle the orthogonality constraint since PCA methods find the basis which represents the maximum variance of data, but not necessarily a physically-meaningful representation of the variables ~\cite{koscielny1982}. Using an oblique rotation onto PCs, the orthogonality problem can be addressed ~\cite{rotatedPCA}. The promax rotation ~\cite{promax} is based on the varimax orthonormal criterion which finds a rotation of a subset of PCs to maximize variance ~\cite{kaiser1958}. %Promax rotation use this new rotated axis and make it oblique maximizing the varimax parameter $Varimax(v_{p}(\sigma))=Varimax(p,\sigma)$.%
Optimizing the varimax parameter will also allow us to find a proper $\sigma$ value and a number of PCs which makes the PCs more physically interpretable.

\subsection{Illustrative example}
\label{subsec:example}

\begin{figure}[ht!]
\begin{center}
\setlength{\tabcolsep}{-10pt} 
\begin{tabular}{ccc}
{\em PC1 (99.97\%)} & {\em PC2 (0.03\%)} \\
\begin{turn}{90} \hspace{0.35cm} \begin{turn}{-90} a) \end{turn} \end{turn}
\includegraphics[width=6.8cm]{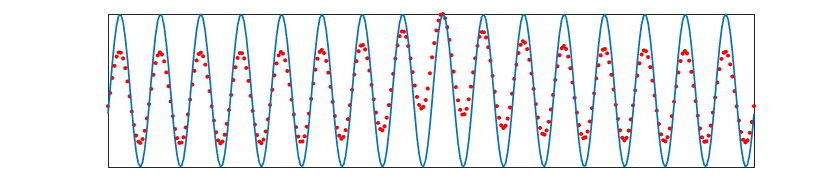}  &
\includegraphics[width=6.8cm]{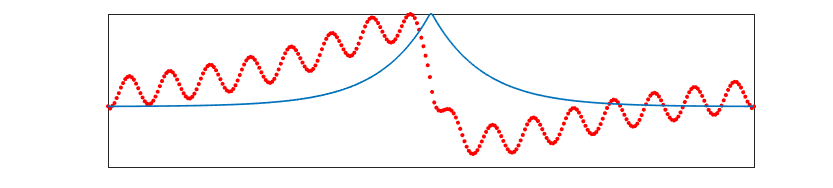}      \\ 
{\em PC1 (44.46\%)} & {\em PC3 (11.69\%)} \\
\begin{turn}{90} \hspace{0.35cm} \begin{turn}{-90} b) \end{turn} \end{turn}
\includegraphics[width=6.8cm]{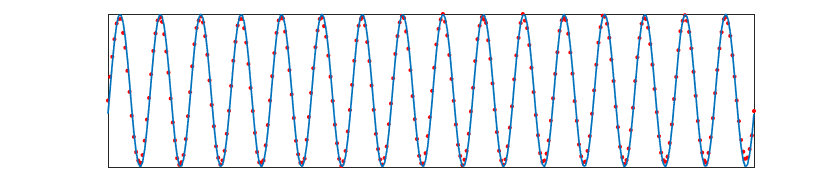}  &
\includegraphics[width=6.8cm]{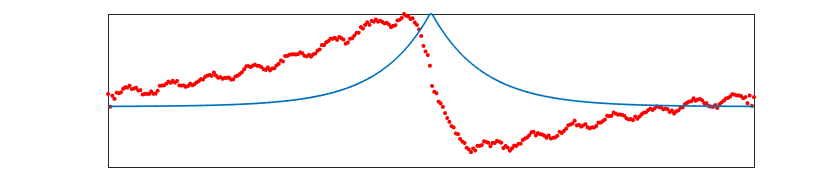}      \\ 
{\em PC1 (99.91\%)} & {\em PC3 (0.03\%)} \\
\begin{turn}{90} \hspace{0.35cm} \begin{turn}{-90} c) \end{turn} \end{turn}
\includegraphics[width=6.8cm]{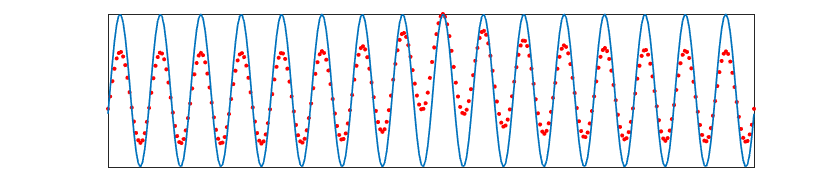}  &
\includegraphics[width=6.8cm]{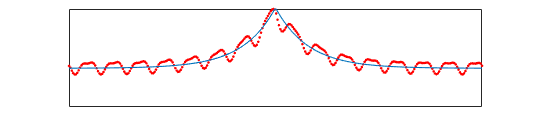}      \\
{\em PC1 (45.87\%)} & {\em PC3 (14.67\%)} \\
\begin{turn}{90} \hspace{0.35cm} \begin{turn}{-90} d) \end{turn} \end{turn}
\includegraphics[width=6.8cm]{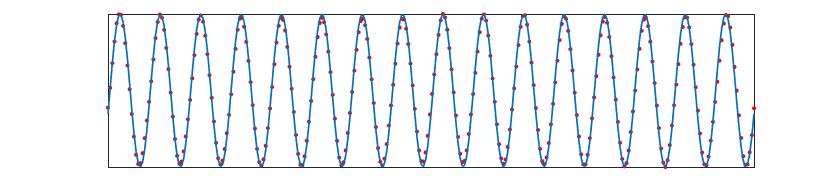}  &
\includegraphics[width=6.8cm]{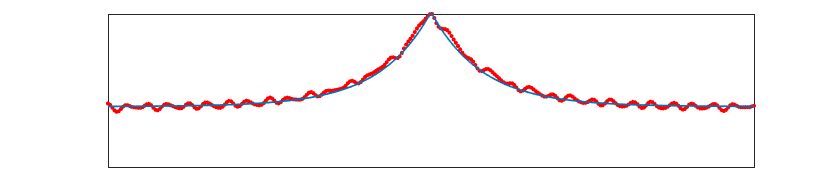}    \\
\end{tabular}
\end{center}
\vspace{-0.8cm}
\caption{\label{fig:PCcompared} Time series: We show the real part from the estimated principal components (red) and his theoretical function (blue) with his explained variance on top. a) Complex PCA and b) Rotated Complex PCA represent the linear methods, and the c) Complex Kernel PCA and d) ROCK-PCA the nonlinear approaches.}
\end{figure}

We illustrate the benefits of the proposed ROCK PCA in a simple toy example. A theoretical spatio-temporal dataset has been generated following the function
$$F(x,y,t)=e^{-|t|}cos(w_{x}x)+sin(w_{t}t)cos(w_{y}y) $$
with two independent time-series and spatial dependence, fixing the time range between $[-2\pi:2\pi]$ with $w_{t}=2\pi f=8$. Optimizing the varimax function, we obtain four PCs and $\sigma=793$, but only two of them represent the theoretical data (first and third in the rotated case). Note that in the case of rotation, one frequently finds two identical PCs, since the oblique rotation allows correlated PCs. As we show in Fig.~\ref{fig:PCcompared} the kernel method find a close decomposition from the theoretical time-series but the rotation is needed to obtain the closest estimation.

%%%%%%%%%%%%%%%%%%%%%%%%
\section{Experimental results}

This section presents the application of the proposed ROCK PCA in the analysis of the spatial and temporal relations in 7 years of Earth Observation (EO) SM data. 

\subsection{SMOS data}
\label{sec:smos}

Global SMOS soil moisture data from the Barcelona Expert Center has been used in this study (http://bec.icm.csic.es/). Since its launch in 2009, SMOS is providing global maps of the Earth's surface soil moisture (top 5 cm) every 3-days with a spatial resolution of $\sim$50 km and a target accuracy of 0.04 m$^3$/m$^3$. For this work, we chose the first seven years of SMOS observations, after its commissioning phase (from May 2010 to May 2017). Note that this period encloses the 2015-2016 El Ni{\~n}o event --the so-called \emph{el Ni{\~n}o Godzilla}, which will be studied.

\subsection{SM spatio-temporal decomposition}

The ROCK PCA analysis has been applied to the SM maps to obtain a multimodal decomposition. The explained variance for the top four components account for $98.2\%$ of the total variance: 70.6 \%(PC1), 13.5\% (PC2), 10.2 \% (PC3) and 4.3\% (PC4). The first component returns an annual oscillation with a $12$-month period and a global spatial distribution, representing the principal SM global trend. The second and fourth components are cast as `resonances' of the annual SM cycle with six- and four-months periods which represent subcycles of global SM as ITCZ (Intertropical Convergence Zona) oscillation (PC2) or seasonal transition (PC4). Interestingly, the third component (PC3) shows a non-seasonal oscillation with a principal period of $4.5$ years trend as ENSO anomalies ($3-7$ years) which can be interpreted as a long-term change in the global SM distribution.

\begin{figure}[ht!]
\begin{center}
\setlength{\tabcolsep}{-8pt} 
\begin{tabular}{cc}
{\em (a) PC1 (70.6)\%} & {\em (b) PC2 (13.5)\%} \\
\includegraphics[width=6cm]{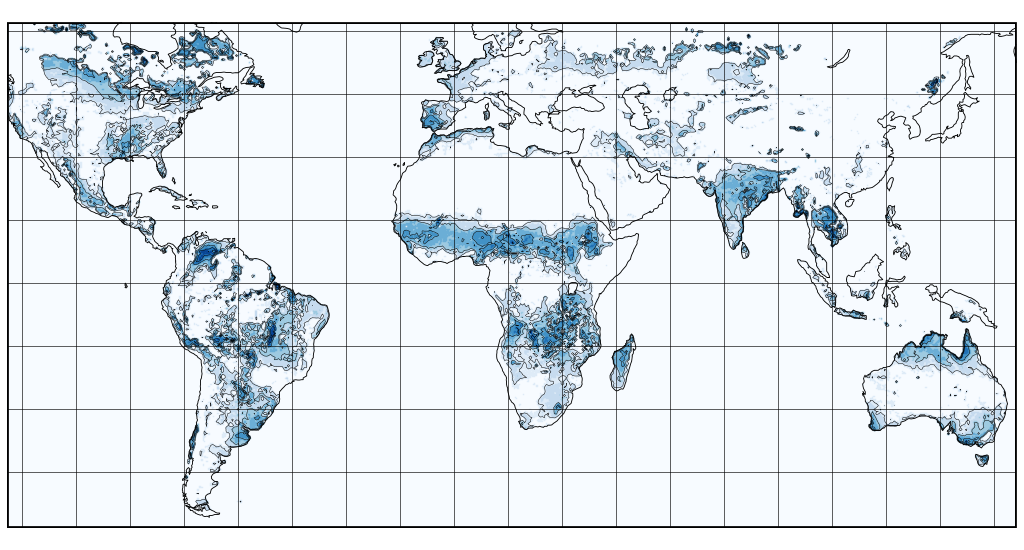}  &
\includegraphics[width=6cm]{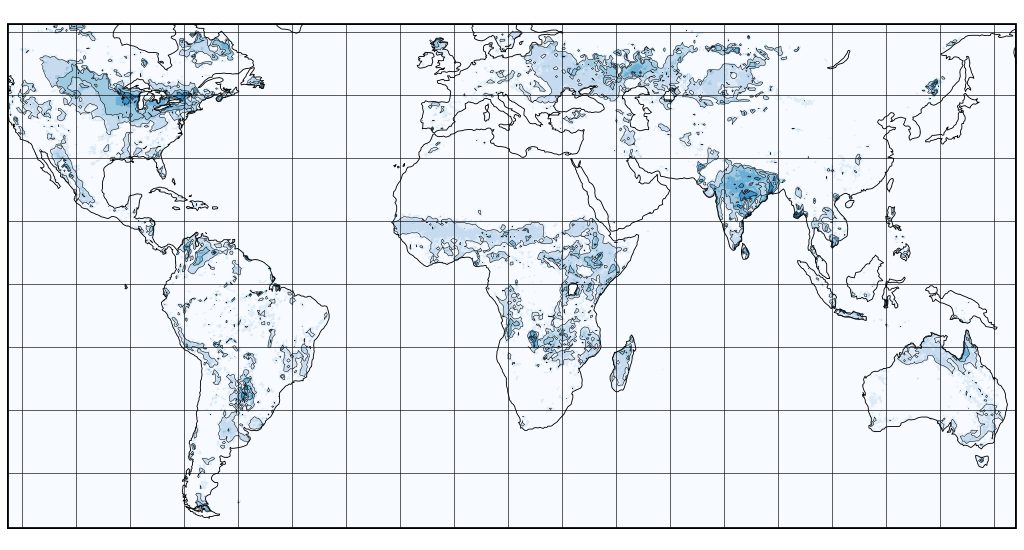}      \\ 
\includegraphics[width=6.8cm]{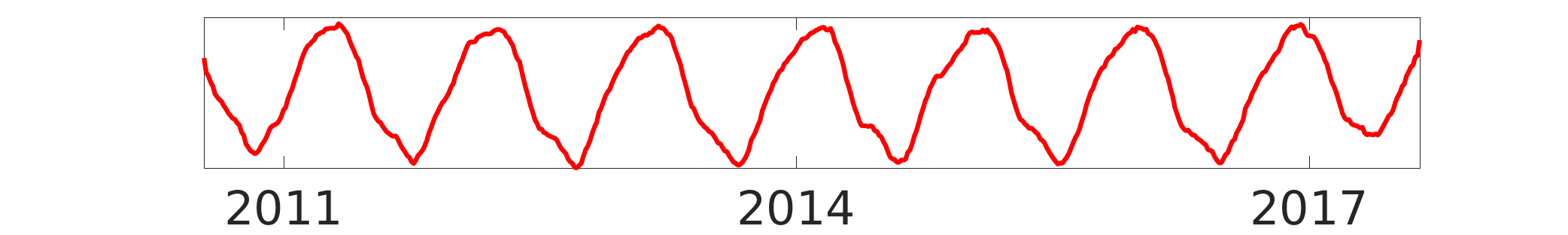}  &
\includegraphics[width=6.8cm]{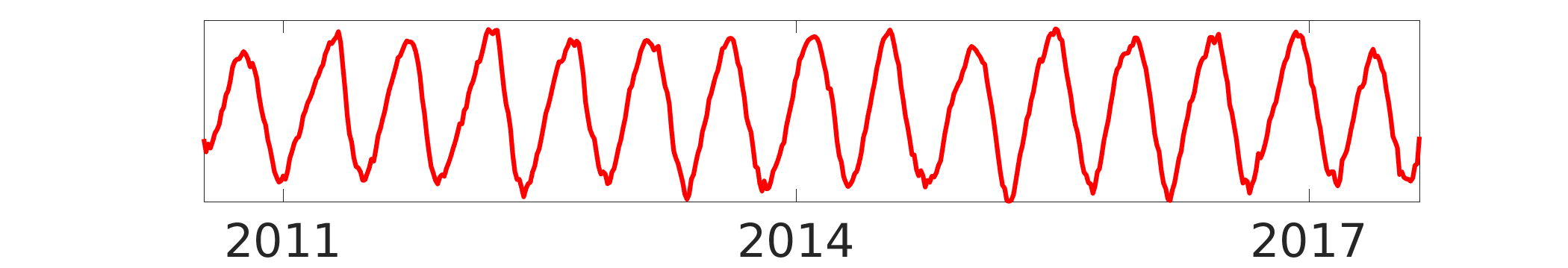}      \\ 
{\em (c) PC3 (10.2)\%} & {\em (d) PC4 (4.3)\%} \\
\includegraphics[width=6cm]{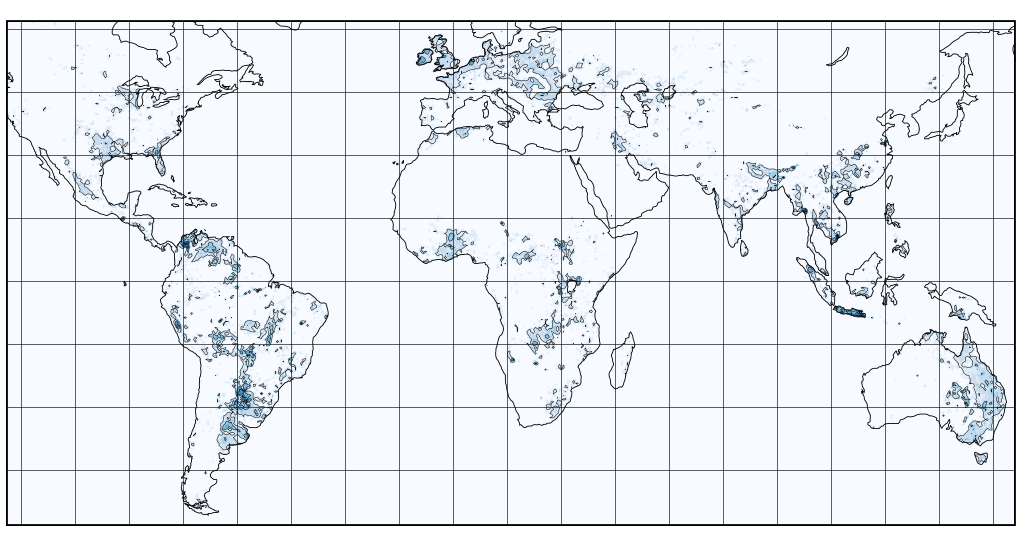}  &
\includegraphics[width=6cm]{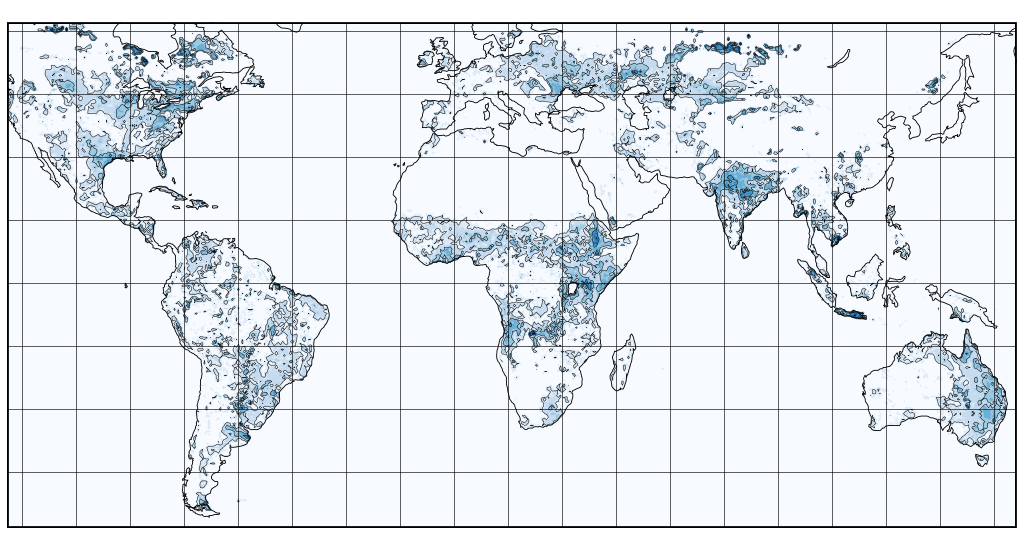} \\
\includegraphics[width=6.8cm]{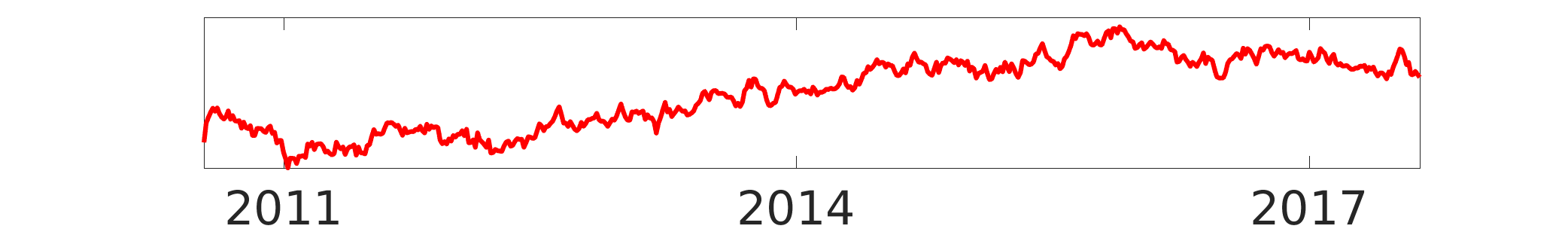}  &
\includegraphics[width=6.8cm]{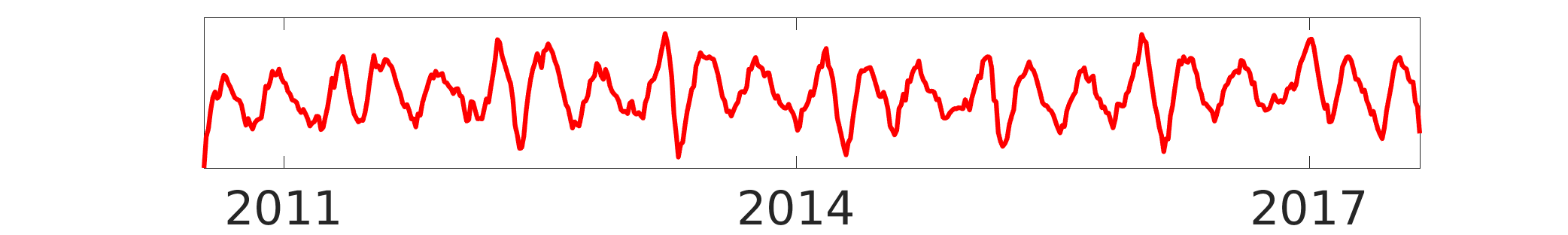} \\
\end{tabular}
\end{center}
\vspace{-1cm}
\end{figure}

\begin{figure}[ht!]
\centerline{\includegraphics[width=0.65\textwidth,scale=0.4]{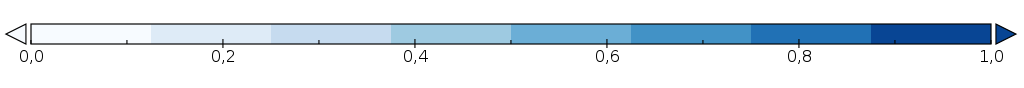}}
\vspace{-0.5cm}
\caption{\label{fig:SMdecomp}Normalized Spatial distribution of amplitude (top) and his respective time series (bottom) for a) PC1, b) PC2, c) PC3 and d) PC4 with his explained variance respectively.}
\end{figure}

Figure~\ref{fig:SMdecomp} shows the (normalized) amplitudes of the spatial maps, which are the absolute value of the spatial projection of the PCs representing the spatial distribution of each time series. PC3 has a restricted spatial distribution with an homogeneous spatial phase that represent SM change which is strongly linked to the ENSO.

\subsection{Connection between the ENSO and the SM trend}

\begin{figure}[ht!]
\begin{tabular}{cc}
a) & b) \\
\includegraphics[width=0.5\textwidth,scale=0.4]{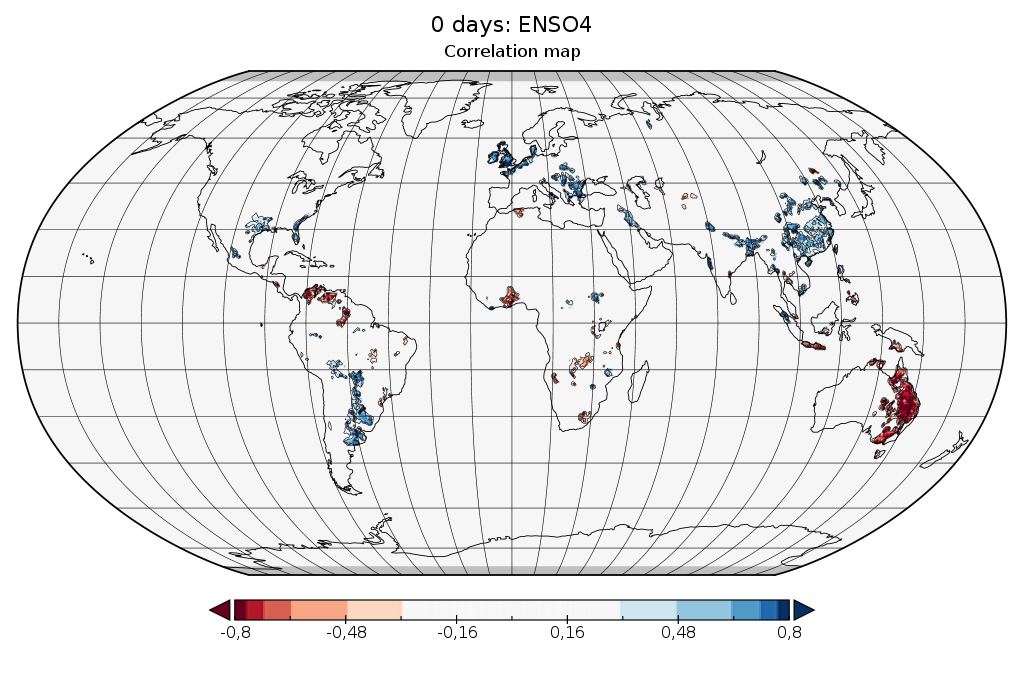} & \includegraphics[width=0.4\textwidth,scale=0.1]{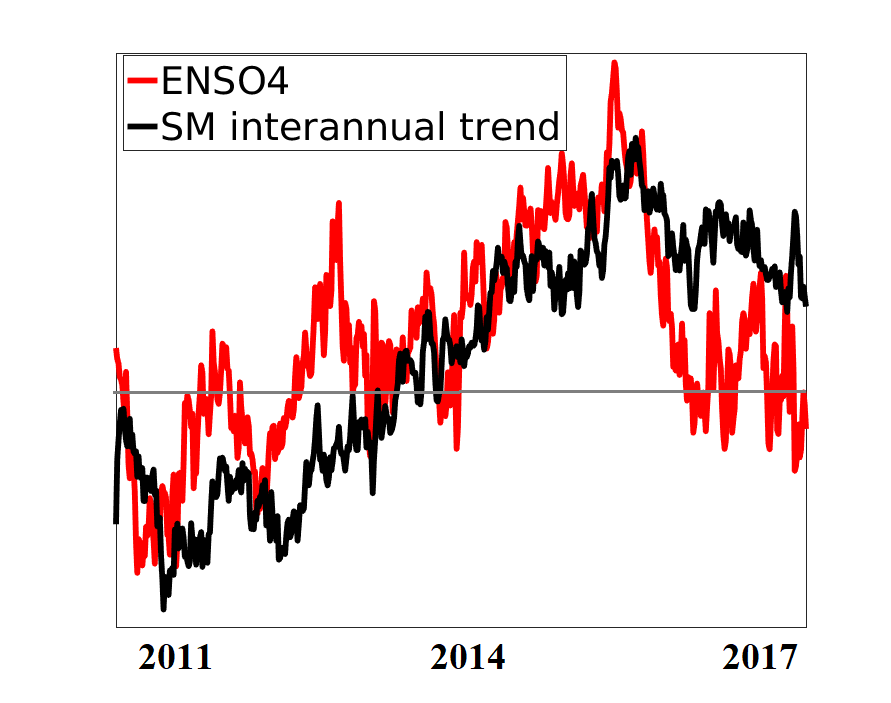} \\
\end{tabular}
\vspace{-0.5cm}
\caption{\label{fig:corr}a) Correlation map between SM long term trend and ENSO4 time series. Spatial dry-wet patterns can be identified. b) figure shows the
temporal evolution of PC3 and the ENSO4 index during the study period.}
\end{figure}

In this analysis, the identified SM cycles and long-term trends have been compared to the ENSO4 index, which represent the SST (Sea Surface Temperature) anomalies from western Pacific Ocean. To do so, we computed the Spearman correlation coefficient between the ENSO4 index and the SM long-trend with the complex spatial phase added to show the spatial connection of the variables. Results are shown in Figure~\ref{fig:corr}. It can be seen that there is a significant association between the SM long-trend and the ENSO4 index with a maximum absolute correlation of 0.8 with a $p$-value below of $10^{-16}$. 

The observed relationship shows the spatial distribution and strength of ENSO influence in wet-dry patterns, as previous studies ~\cite{miralles2013} but using only direct EO data and also a wider spatial wet-dry pattern. Figure~\ref{fig:corr} presents the spatial distribution of positive and negative correlation coefficients: an increasing SM trend can be observed for instance in eastern Europe, southeastern South America and Southern US whereas a decreasing SM trend is visible in Australia's eastern cost, northern South America, northern gulf of Guinea. This connection shows an intuitive and direct relation between the ENSO4 dynamics and the extracted SM long term trends. 

%%%%%%%%%%%%%%%%%%%%%%%%%%%%%%%%%%%%%%%%%%%%%%%%%%%%%%%%%%%%%%%%%%%%%%%%%%%%%%%%%%%%%%%%%%%%%%%%%%

\section{Conclusion}

In this paper we propose the ROCK PCA method for spatial-temporal analysis and its improvement compared to classical PCA methods. Our approach unveiled both time and space modes, trends and periodicities unlike standard PCA decompositions. We illustrated the performance with a toy example with nonlinear time and space feature relations showing his ability for nonlinear data decomposition. 
we then showed a preliminary study of its applications to global satellite SM estimates that allowed associate its long-term component to ENSO. This method is a powerful tool for the study of global dynamics of EO data and can ultimately bring more information about climate teleconnections.

%%%%%%%%%%%%%%%%%%%%%%%%%%%%%%%%%%%%%%%%%%%%%%%%%%%%%%%%%%%%%%%%%%%%%%%%%%%%%%%%%%%%%%%%%%%%%%%%%%

\small
\bibliographystyle{IEEEbib}
\bibliography{biblio}

\end{document}